\documentclass[english,aps,pra,twocolumn,floatfix]{revtex4}

\usepackage{amsmath}
\usepackage{amssymb}
\usepackage{amsfonts}
\usepackage{bm}
\usepackage{bbm}
\usepackage{epsfig}
\usepackage{grffile}
\usepackage{babel}

\usepackage[usenames,dvipsnames]{color}
\definecolor{grey}{rgb}{.5,.5,.5}
\definecolor{dblue}{rgb}{0,0,.5}
\definecolor{dgreen}{rgb}{0,.65,0}

\usepackage[colorlinks=true,citecolor=dblue,linkcolor=dblue]{hyperref}
\usepackage[all]{hypcap}

\newcommand{\id}{\mathbbm{1}}
\newcommand{\Tr}{\operatorname{Tr}}
\newcommand{\aux}{{\mathrm{aux}}}
\newcommand{\tot}{{\mathrm{tot}}}
\newcommand{\vac}{{\mathrm{vac}}}
\newcommand{\diag}{{\mathrm{diag}}}
\newcommand{\Span}{{\mathrm{span}}}
\newcommand{\bra}{\langle}
\newcommand{\ket}{\rangle}
\newcommand{\mc}[1]{\mathcal{#1}}

\newcommand{\ce}{{\mathrm{c}}}
\newcommand{\gce}{{\mathrm{gc}}}

\renewcommand{\H}{\mc{H}}
\newcommand{\hH}{\hat{H}}
\newcommand{\hS}{\hat{S}}
\newcommand{\hq}{\hat{q}}
\newcommand{\hQ}{\hat{Q}}
\newcommand{\hX}{\hat{X}}
\newcommand{\hvS}{\hat{\vec{S}}}
\newcommand{\dm}{{\hat{\varrho}}}

\newcommand{\dmp}{\varrho}
\newcommand{\epsP}{\epsilon_{\mathrm{P}}}

\renewcommand{\vec}[1]{{\boldsymbol{#1}}}

\newcommand{\vn}{{\vec{n}}}
\newcommand{\ua}{{\uparrow}}
\newcommand{\da}{{\downarrow}}

\newcommand{\duke} {Department of Physics, Duke University, Durham, North Carolina 27708, USA}

\begin{document}

\newcommand{\titlefont}{\fontfamily{ptm}\selectfont}
\title{\titlefont Matrix product purifications for canonical ensembles and quantum number distributions}
\author{Thomas Barthel}
\affiliation{\duke}
\date{June 06, 2016}

\begin{abstract}
Matrix product purifications (MPPs) are a very efficient tool for the simulation of strongly correlated quantum many-body systems at finite temperatures. When a system features symmetries, these can be used to reduce computation costs substantially. It is straightforward to compute an MPP of a grand-canonical ensemble, also when symmetries are exploited. This paper provides and demonstrates methods for the efficient computation of MPPs of canonical ensembles under utilization of symmetries. Furthermore, we present a scheme for the evaluation of global quantum number distributions using matrix product density operators (MPDOs). We provide exact matrix product representations for canonical infinite-temperature states, and discuss how they can be constructed alternatively by applying matrix product operators to vacuum-type states or by using entangler Hamiltonians.
A demonstration of the techniques for Heisenberg spin-$1/2$ chains explains why the difference in the energy densities of canonical and grand-canonical ensembles decays as $1/L$.
\end{abstract}


\maketitle

\section{Introduction}
Strongly correlated one-dimensional (1D) and quasi-1D systems can be simulated accurately using the density matrix renormalization group (DMRG) \cite{White1992-11,White1993-10,Schollwoeck2005}, which is a set of algorithms operating on matrix product states (MPS) \cite{Fannes1992-144,Rommer1997,Schollwoeck2011-326}. Real systems can usually not be isolated completely from their environment and experiments are hence necessarily done at finite (nonzero) temperatures. To allow for a direct investigation of experimental results or for theoretical investigations on the influence of temperature, DMRG techniques have been generalized to describe thermal states. 

The historically first DMRG algorithm for finite temperatures is the quantum transfer-matrix renormalization group \cite{Nishino1995-64,Bursill1996-8,Shibata1997-66,Wang1997-56}, which is, however, complicated by the non-hermiticity of the quantum transfer matrix and does not allow for a direct evaluation of non-local observables. A recent approach combines time-dependent DMRG \cite{Vidal2003-10,White2004,Daley2004} with Monte Carlo to evaluate thermal observables by sampling over so-called minimally entangled typical thermal states \cite{White2009-102} and is also applicable for the study response functions \cite{Binder2015-92,Bruognolo2015-92}. 

The arguably most successful method, which is the focus of this paper, is based on a purification of the density matrix \cite{Uhlmann1976,Uhlmann1986,Nielsen2000}, i.e., the representation through a pure state in an enlarged Hilbert space \footnote{A state $|\phi_\dmp\ket\in\H\otimes\H_\aux$ is called a purification of the density matrix $\dm$ on $\H$ if $\Tr_\aux |\phi_\dmp\ket\bra \phi_\dmp|=\dm$.}. For numerical purposes, these are approximated very precisely in matrix product form, to obtain so-called matrix product purifications (MPPs) \cite{Verstraete2004-6}. First applications to static phenomena \cite{Feiguin2005-72,Barthel2005} showed the potential of the approach which is also known as the ancilla method. It has been extended for the computation of response functions \cite{Barthel2009-79b,Karrasch2012-108,Barthel2012_12,Barthel2013-15} and allows it also to work directly in the frequency domain \cite{Tiegel2014-90,Holzner2011-83}. The various applications comprise, for example, the phase diagrams of spin-ladder systems \cite{Ruegg2008-101,Bouillot2011-83}, thermometry for ultracold bosons in optical lattices \cite{Gori2016-93}, the Drude weight in spin-$1/2$ and Hubbard chains \cite{Karrasch2012-108,Karrasch2013-87,Karrasch2014-90}, as well as spectral properties of quantum magnets \cite{Lake2013-111,Tiegel2016-93}, bosonic systems \cite{Barthel2012_12}, and spin-incoherent Luttinger liquids \cite{Feiguin2010-81}.

It is straightforward to compute an MPP of a grand-canonical ensemble. We consider lattice systems (1D or quasi-1D), such that the Hilbert space has tensor-product structure. The grand-canonical infinite-temperature state is then a tensor product of on-site identities and the corresponding MPP has bond dimension one. A subsequent imaginary-time evolution yields the MPP of the grand-canonical ensemble at finite temperatures. When a system features symmetries, these can be exploited to reduce computation costs substantially. For a conserved quantity $\hat{Q}$, one can use that the imaginary-time evolution does not mix components with different $\hat{Q}$ quantum numbers. However, the grand-canonical ensemble contains, of course, contributions from all quantum number sectors such that the benefit of symmetries is reduced.

This paper introduces and demonstrates efficient methods to compute MPPs of canonical ensembles under utilization of symmetries. In particular, we provide exact matrix product representations for canonical infinite-temperature states (Sec.~\ref{sec:CE_beta0_exact}) and discuss how they can be constructed alternatively by applying matrix product operators to vacuum-type states (Sec.~\ref{sec:CE_beta0_MPO}). We also discuss an earlier variational approach using entangler Hamiltonians \cite{Feiguin2010-81,Nocera2016-93} and present a generalized version of it (Sec.~\ref{sec:entangler}).
Furthermore, we introduce a method for the evaluation of global quantum number probability distributions using matrix product density operators (MPDOs) \cite{Verstraete2004-6,Zwolak2004-93} (Sec.~\ref{sec:MPDO_QN-distr}). The techniques are exemplified for antiferromagnetic spin-$1/2$ Heisenberg chains, explaining why the difference in the energy densities of canonical and grand-canonical ensembles with the same magnetization decays as $1/L$ with the system size $L$ (Sec.~\ref{sec:example}).

\section{Symmetries in matrix products} \label{sec:symmetries}
For simplicity, let us restrict our considerations to an Abelian symmetry with a single conserved quantity $\hQ$ such as the total number of particles or magnetization ($[\hH,\hQ]=0$). However, everything generalizes in a very similar manner to the cases with multiple conserved quantities and non-Abelian symmetries \cite{McCulloch2001-5,Weichselbaum2012-327}. For the latter one exploits that dependencies inside each multiplet are given by Clebsch-Gordan coefficients as exemplified by the Wigner-Eckart theorem.

Let us consider a lattice system with $L$ sites, orthonormal on-site basis states $\{|n_i\ket\,|\,n_i=0,\dotsc,d-1\}$, and associated quantum numbers $q(n_i)$, i.e., $\hq_i|n_i\ket=q(n_i)|n_i\ket$, where $\hQ:=\sum_{i=1}^L\hq_i$ is the conserved quantity. A \emph{matrix product state} has the form
\begin{equation}\label{eq:MPS}
	|\psi\ket = \sum_{\vn} A^{n_1}_1 A^{n_2}_2\dotsb A^{n_L}_L|\vn\ket,
\end{equation}
with $\vn:=(n_1,\dotsc,n_L)$ and $D_{i-1}\times D_{i}$ matrices $A^{n_i}_i$. The $D_i$ are referred to as \emph{bond dimensions}. For the matrix product to yield a scalar, we require $D_0=D_{L}=1$.

One can enforce the MPS $|\psi\ket$ to have quantum number $Q$ by assigning quantum numbers to the bond indices $a_i$ and $b_i$ of the matrices, imposing that matrix elements $[A^{n_i}_i]_{a_i,b_i}$ can be nonzero only if the quantum number constraint
\begin{equation}\label{eq:qnConstraint}
	q(b_i)=q(a_i)+q(n_i)
\end{equation}
is obeyed, and enforcing $q(a_1)=0$ and $q(b_L)=Q$. As in the matrix product $b_i=a_{i+1}$, the quantum number constraint leads indeed to
\begin{align*}
	\sum_{i=1}^L q(n_i)
	 &= -q(a_1)+q(b_1)+\sum_{i=2}^L q(n_i)\\
	 &= -q(a_1)+q(b_2)+\sum_{i=3}^L q(n_i)\\
	 &= \dotso = -q(a_1)+q(b_L) = Q
\end{align*}
for all nonzero contributions in the MPS \eqref{eq:MPS}, such that $\hQ|\psi\ket=Q|\psi\ket$.

This is in agreement with the traditional interpretation of the MPS tensors $A_i$ in the DMRG \cite{Rommer1997}. From that point of view, the tensors take the role of projectors from a reduced Hilbert space of dimension $d D_{i-1}$ with basis $|a_i\ket\otimes|n_i\ket$ for the block of sites $[1,i]$ to another reduced $D_i$-dimensional Hilbert space with basis $|b_i\ket$ for the same block, such that $|b_i\ket=\sum_{a_i,n_i}[A^{n_i}_i]_{a_i,b_i}|a_i\ket\otimes|n_i\ket$. The constraint \eqref{eq:qnConstraint} then means that mixing of quantum numbers is prohibited in these projections.

For systems with multiple conserved quantities, $\hQ^{(j)}=\sum_{i=1}^L\hq^{(j)}_i$, the only difference to the above is that quantum number labels become vectors such that $\vec{q}(b_i)=\vec{q}(a_i)+\vec{q}(n_i)$, $\vec{q}(a_1)=(0,\dotsc,0)$ and $\vec{q}(b_L)=\vec{Q}$.

The explicit implementation of symmetries can reduce computation costs considerably. Typical costly operations for matrix product states \eqref{eq:MPS}, e.g., in a time-evolution or ground-state calculation, are singular value decompositions of the $A$-tensors like $A_i^{n_i}=U \Lambda V^{n_i}$, where $U$ and $V$ are isometric according to $U^\dag U=\id$ and $\sum_{n_i}V^{n_i}(V^{n_i})^\dag=\id$, and $\Lambda$ is a diagonal matrix containing the singular values. The cost of such a singular value decomposition scales as $\mc O(dD^3)$ with the on-site Hilbert space dimension $d$ and the bond dimension $D$. Now, with the implementation of symmetries according to the quantum number constraint, the $A$-tensors assume a block structure with nonzero blocks corresponding to groups of states obeying Eq.~\eqref{eq:qnConstraint}. A singular value decomposition can then be done block by block. Often, the total computation cost is then dominated by the biggest block, the dimensions of which can be much smaller than the total bond dimensions $D_i$.

\section{Considered thermal ensembles} \label{sec:ensembles}
The Hamiltonian $\hH$ commutes with the conserved quantity $\hQ$. In a situation where the system exchanges energy with a bath but these interactions commute with $\hQ$, according to Jaynes' maximum entropy principle from statistical mechanics \cite{Jaynes1957-106}, the equilibrium state of the system is given by the (here, unnormalized) \emph{canonical ensemble}
\begin{equation}\label{eq:CE}
	\dm^\ce_{\beta,Q}:= e^{-\beta\hH_Q}\quad\text{on $\H_Q$},
\end{equation}
with $\hH_Q$ being the component of the Hamiltonian in the quantum number $Q$ subspace $\H_Q$ of the full Hilbert space $\H=\bigoplus_Q \H_Q$.
According to the same principle, the equilibrium state is the \emph{grand-canonical ensemble}
\begin{equation}\label{eq:GCE}
	\dm^\gce_{\beta,\alpha}= e^{-\beta(\hH+\alpha\hQ)}
\end{equation}
if system and bath also exchange the quantity associated with $\hQ$. Here, $\alpha$ is the associated Lagrange multiplier that fixes the expectation value of $\hQ$. Two examples are that $\hQ$ is the total magnetization $\sum_i\hS^z_i$ and $-\alpha$ the magnetic field, or that $\hQ$ is the total particle number $\sum_i\hat{n}_i$ and $-\alpha$ the chemical potential.

In more complex cases with multiple conserved quantities $\hQ^{(j)}$, one can also consider ensembles like $\exp[{-\beta(\hH_{Q^{(1)}}+\alpha_2\hQ^{(2)})}]$, where $\hQ^{(1)}$ is fixed to $Q^{(1)}$ and the expectation value $\bra\hQ^{(2)}\ket$ is fixed by the Lagrange multiplier $\alpha_2$. An example is the Fermi-Hubbard model, where the local basis states for site $i$ are $|n_{i,\ua},n_{i,\da}\ket$ with electron spin $\sigma\in\{\ua,\da\}$ and particle number $n_{i,\sigma}\in\{0,1\}$. The total number of electrons $\hat{N}=\sum_{i,\sigma}\hat{n}_{i,\sigma}$ could be fixed, but, for example due to a transverse magnetic field, only the expectation value of the magnetization $\bra \hS^z_\tot\ket=\frac{1}{2}\sum_i \bra \hat{n}_{i,\ua}-\hat{n}_{i,\da}\ket$ might be fixed by a magnetic field in $z$-direction.
In passing, we will also comment on such ensembles.

\section{Matrix product representations of grand-canonical ensembles} \label{sec:GCE}
We want to compute finite-temperature expectation values $\bra\hat{O}\ket=\Tr(\dm\,\hat{O})/Z$ with respect to the grand-canonical ensemble \eqref{eq:GCE} using matrix product purifications and exploiting the $\hQ$-conservation. The partition function $Z\equiv \Tr \dm$ is needed for normalization. As a purification of the density matrix $\dm^\gce_{\beta,\alpha}$ on $\H$, we can employ the vectorization $|\dmp^\gce_{\beta/2,\alpha}\ket\in \H\otimes\H$ of the density matrix at twice the temperature \cite{Binder2015-92}. The \emph{vectorization} $|Y\ket\in\H\otimes\H$ of any operator $\hat{Y}$ on $\H$ with respect to the orthonormal basis $\{|\vn\ket\}$ is defined as
\begin{equation*}
	|Y\ket \equiv \sum_{\vn,\vn'}\bra\vn|\hat{Y}|\vn'\ket \,|\vn\ket\otimes|\vn'\ket.
\end{equation*}
The vectorization $|\dmp^\gce_{\beta/2,\alpha}\ket$ of $\dm^\gce_{\beta/2,\alpha}$ is a purification of the density matrix $\dm^\gce_{\beta,\alpha}$ in the sense that
\begin{equation*}
	\Tr_\aux |\dmp^\gce_{\beta/2,\alpha}\ket\bra \dmp^\gce_{\beta/2,\alpha}|=\big(\dm^\gce_{\beta/2,\alpha}\big)^2=\dm^\gce_{\beta,\alpha},
\end{equation*}
where the partial trace is taken over the second (\emph{auxiliary} or \emph{ancilla}) part of the tensor product space $\H\otimes\H$. With the purification, one can compute expectation values as
\begin{equation}\label{eq:GCE_expect}
	\bra\hat{O}\ket^\gce_{\beta,\alpha} = \frac{\bra\dmp^\gce_{\beta/2,\alpha}|\hat{O}|\dmp^\gce_{\beta/2,\alpha}\ket}{\bra\dmp^\gce_{\beta/2,\alpha}|\dmp^\gce_{\beta/2,\alpha}\ket}.
\end{equation}

An accurate matrix product approximation of the purification in the form
\begin{equation}\label{eq:MPP}
	|\dmp\ket=\sum_{\vn,\vn'} A^{n_1,n_1'}_1 A^{n_2,n_2'}_2\dotsb A^{n_L,n_L'}_L|\vn\ket\otimes|\vn'\ket
\end{equation}
can be constructed by starting at infinite temperature ($\beta=0$) with
\begin{subequations}\label{eq:GCE_beta0}
\begin{align}
	\dm^\gce_{0,\alpha}&= \id =\bigotimes_i \big(\sum_{n_i}|n_i\ket\bra n_i|\big)\\ \label{eq:MPP_GCE_beta0}
	\quad\Leftrightarrow\quad
	|\dmp^\gce_{0,\alpha}\ket&=\bigotimes_i \big(\sum_{n_i}|n_i\ket\otimes|n_i\ket_\aux\big),
\end{align}
\end{subequations}
where $|n_i\ket_\aux$ is a basis state for site $i$ of the second (auxiliary) part in $\H\otimes\H$.
An exact matrix product representation \eqref{eq:MPP} with bond dimensions $D_i=1$ is given by $[A^{n_i,n_i'}_i]_{1,1}=\delta_{n_i,n_i'}$. With $|\dmp^\gce_{0,\alpha}\ket$ as the initial state, one can time evolve the MPP \eqref{eq:MPP} in imaginary time using the time-dependent DMRG algorithm (tDMRG) \cite{White2004,Daley2004} or time-evolved block decimation (TEBD) \cite{Vidal2003-10} to obtain an MPP for the finite-temperature state $\dm^\gce_{\beta,\alpha}$,
\begin{equation}\label{eq:MPP_GCE_beta}
	|\dmp^\gce_{\beta/2,\alpha}\ket = \big(e^{-\beta(\hH+\alpha\hQ)/2} \otimes \id\big) |\dmp^\gce_{0,\alpha}\ket.
\end{equation}
As the purification $|\dmp^\gce_{\beta/2,\alpha}\ket$ of $\dm^\gce_{\beta,\alpha}$ is simply the vectorization of $\dm^\gce_{\beta/2,\alpha}$, Eq.~\eqref{eq:MPP_GCE_beta} yields at the same time an approximation of $\dm^\gce_{\beta/2,\alpha}$ in MPDO form \cite{Barthel2013-15}.

While the evolution operator in Eq.~\eqref{eq:MPP_GCE_beta} commutes with both $\hQ\otimes \id$ and $\id\otimes\hQ$, we cannot utilize the corresponding conservation laws, because the initial state $|\dmp^\gce_{0,\alpha}\ket$ is not an eigenstate of either of them. All quantum number sectors contribute to the infinite-temperature state \eqref{eq:GCE_beta0}. However, it is an eigenstate of
\begin{equation}
	\hat{\mc{Q}} := \hQ\otimes \id - \id\otimes\hQ
\end{equation}
with eigenvalue $\mc{Q}=0$.
In analogy to what was described in section~\ref{sec:symmetries}, this can be exploited to reduce computation costs. With local quantum numbers $\mathfrak{q}(n_i,n_i')=q(n_i)-q(n_i')$, the quantum number constraint for the tensors $[A^{n_i,n_i'}_i]_{a_i,b_i}$ in the matrix product \eqref{eq:MPP} reads $\mathfrak{q}(b_i)=\mathfrak{q}(a_i)+\mathfrak{q}(n_i,n_i')$, in analogy to Eq.~\eqref{eq:qnConstraint}. With $\mathfrak{q}(a_1)=\mathfrak{q}(b_L)=0$, it guarantees that $\sum_i\mathfrak{q}(n_i,n_i')={\mc{Q}}=0$ is conserved in all operations.

\section{Matrix product representations of canonical ensembles} \label{sec:CE}
An MPP or MPDO for the canonical ensemble \eqref{eq:CE} can be computed quite similarly -- the biggest difference being the initial state at $\beta=0$. The restriction of the identity to the quantum number $Q$ subspace $\H_Q$ can be written as
\begin{align}\nonumber
	\dm^\ce_{0,Q} &= \id_Q = \delta_{\hQ,Q} = \sum_{\vn,\sum_i q(n_i)=Q} |\vn\ket\bra\vn|\\ \label{eq:CE_beta0}
	\Leftrightarrow\quad
	|\dmp^\ce_{0,Q}\ket &= \sum_{\vn,\sum_i q(n_i)=Q} |\vn\ket\otimes|\vn\ket.
\end{align}
Clearly, this is an eigenstate of both $\hQ\otimes\id$ and $\id\otimes\hQ$ such that $|\dmp^\ce_{0,Q}\ket\in \H_Q\otimes\H_Q$. In a matrix product representation \eqref{eq:MPP} of it, we can hence exploit the two conservation laws by assigning tuples $(q,\tilde{q})$ of quantum numbers to the bond indices. For bond $(i,i+1)$, these are the eigenvalues of $\sum_{j=1}^i\hq_j\otimes \id$ and $\sum_{j=1}^i\id\otimes\hq_j$, respectively. The quantum number constraints for the tensors $[A^{n_i,n_i'}_i]_{a_i,b_i}$ in the MPP then read
\begin{subequations}
\begin{align}
	q(b_i)&=q(a_i)+q(n_i) \quad\text{and}\\
	\tilde{q}(b_i)&=\tilde{q}(a_i)+q(n_i'),
\end{align}
\end{subequations}
in analogy to Eq.~\eqref{eq:qnConstraint}. With $q(a_1)=\tilde{q}(a_1)=0$ and $q(b_L)=\tilde{q}(b_L)=Q$, it is guaranteed that $\sum_i q(n_i)=\sum_i q(n'_i)=Q$ is conserved in all operations.

MPS methods to construct the infinite-temperature state are described in the next section. In the subsequent imaginary-time evolution
\begin{equation*}
	|\dmp^\ce_{\beta/2,Q}\ket = \big(e^{-\beta\hH/2} \otimes \id\big) |\dmp^\ce_{0,Q}\ket,
\end{equation*}
which can be implemented using tDMRG, the growth of the MPS bond dimensions $D_i$ is controlled by truncations. Details on the truncation scheme that we employ can be found in Appendix A of Ref.~\cite{Binder2015-92}. Finally, in analogy to the grand-canonical case \eqref{eq:GCE_expect}, one can compute thermal expectation values with the formula 
\begin{equation}
	\bra\hat{O}\ket^\ce_{\beta,Q} = \frac{\bra\dmp^\ce_{\beta/2,Q}|\hat{O}|\dmp^\ce_{\beta/2,Q}\ket}{\bra\dmp^\ce_{\beta/2,Q}|\dmp^\ce_{\beta/2,Q}\ket}.
\end{equation}

\section{Constructing the canonical infinite-temperature states} \label{sec:CE_beta0}
An explicit matrix product representation \eqref{eq:MPP} of the canonical infinite-temperature state \eqref{eq:CE_beta0} is provided in section~\ref{sec:CE_beta0_exact}, but it may be inconvenient to incorporate in existing MPS codes. Therefore, let us first describe an alternative scheme.

\subsection{Applying MPOs to vacuum states}\label{sec:CE_beta0_MPO}
The state \eqref{eq:CE_beta0} can be generated by starting from a projection onto the minimum quantum number state (vacuum) and to then repeatedly apply a matrix product (super-)operator that increases the quantum number until reaching quantum number $Q$ and the state \eqref{eq:CE_beta0}. Let us shortly describe the procedure for systems of bosons, spins, and fermions.

For bosons, let us consider the case where the local bases have been truncated to allow for at most $d-1$ bosons per site. The (projected) ladder operators are
\begin{align*}
	\hat{b}^\dag_i|n_i\ket&=(1-\delta_{n_i,d-1})\sqrt{n_i+1}\,|n_i+1\ket,\quad\text{and}\\
	\hat{b}_i|n_i\ket&=\sqrt{n_i}\,|n_i-1\ket,
\end{align*}
and hence, $\hat{n}_i=\hat{b}^\dag_i\hat{b}_i=\sum_{n_i=0}^{d-1} n_i|n_i\ket\bra n_i|$.
The purification $|\dmp^\ce_{0,N}\ket$ of the identity in the $N$-particle subspace can be written as
\begin{equation}\label{eq:CE_boson}
	|\dmp^\ce_{0,N}\ket = \sum_{\vn,\sum_i n_i=N} \hspace{-2ex}|\vn\ket\otimes|\vn\ket 
	 \propto \frac{\big(\hat{\mc{B}}^\dag_\tot\big)^N}{N!} |\vac\ket\otimes|\vac\ket,
\end{equation}
where $|\vac\ket$ is the vacuum state with $n_i=0$ $\forall_i$ and $\hat{\mc{B}}^\dag_\tot:=\sum_i\hat{b}^\dag_i\otimes \hat{b}^\dag_i$ with the first $\hat{b}^\dag_i$ acting in the primary Hilbert space and the second $\hat{b}^\dag_i$ acting in the auxiliary Hilbert space \footnote{If one does not restrict the maximum number of bosons per site ($d\to\infty$), the proportionality in Eq.~\eqref{eq:CE_boson} becomes an equality. Eq.~\eqref{eq:CE_boson} follows from the multinomial theorem as $(\hat{\mc{B}}^\dag_\tot)^N/N!=\sum_{\vn,\sum n_i=N}\prod_{i=1}^L (\hat{b}^\dag_i)^{n_i}/n_i!$ and $(\hat{b}^\dag\otimes \hat{b}^\dag)^n|0\ket\otimes|0\ket=n!\,|n\ket\otimes|n\ket$}. To this purpose, one first builds an MPS representation \eqref{eq:MPP} of the vacuum state $|\vac\ket\otimes|\vac\ket$ by choosing $1\times 1$ matrices $A^{n_i,n_i'}_i=\delta_{n_i,0}\delta_{n_i',0}$. Alternatively, it can also be generated variationally as the ground state of the Hamiltonian $\sum_i \hat{n}_i\otimes \hat{n}_i$. The operator $\hat{\mc{B}}^\dag_\tot$ can be written as an MPO with bond dimension 2;
\begin{multline*}
	\hat{\mc{B}}^\dag_\tot=
	\begin{bmatrix}\id&\hat{b}^\dag_1\otimes \hat{b}^\dag_1\end{bmatrix}
	\begin{bmatrix}\id&\hat{b}^\dag_2\otimes \hat{b}^\dag_2\\0&\id\end{bmatrix}
	\begin{bmatrix}\id&\hat{b}^\dag_3\otimes \hat{b}^\dag_3\\0&\id\end{bmatrix}\dotsb\\\times
	\begin{bmatrix}\id&\hat{b}^\dag_{L-1}\otimes \hat{b}^\dag_{L-1}\\0&\id\end{bmatrix}
	\begin{bmatrix}\hat{b}^\dag_L\otimes \hat{b}^\dag_L\\\id\end{bmatrix}.
\end{multline*}
According to \eqref{eq:CE_boson}, we apply this MPO $N$ times to the MPS representation of the vacuum state \cite{Schollwoeck2011-326}.

For a spin-$s$ system with local bases $\{|{-s}\ket,\dotsc,|s\ket\}$, one can proceed in exactly the same way, starting from the fully polarized state $|{-s},-s,\dotsc\ket\otimes|{-s},-s,\dotsc\ket$, i.e., the state with eigenvalues $-s$ for all operators $\hS^z_i\otimes \id$ and $\id\otimes \hS^z_i$. In order to obtain an equal weight superposition \eqref{eq:CE_beta0} in the sector with total magnetization $S^z_\tot=-sL+N$, for general $s$, one should not, however, apply ($N$ times) the operator $\hat{\mc{S}}^+_\tot:=\sum_i\hS^+_i\otimes\hS^+_i$ but the analog of the bosonic operator $\hat{\mc{B}}^\dag_\tot=\sum_i\hat{b}^\dag_i\otimes \hat{b}^\dag_i$. To this purpose, one can define the ladder operators as $\hat{b}^\dag_i|{-s}+n_i\ket:=(1-\delta_{n_i,2s})\sqrt{n_i+1}\,|{-s}+n_i+1\ket$. For spin-$1/2$ systems, having single-site Hilbert space dimensions $d=2$, $\hat{\mc{B}}^\dag_\tot$ coincides with $\hat{\mc{S}}^\dag_\tot$.

For fermionic systems, the approach works just as well by replacing the bosonic ladder operators $\hat{b}^\dag_i$ with fermionic ladder operators $\hat{c}^\dag_i$. In the construction of the infinite-temperature purification, one does not need to worry about fermionic sign factors, as operators always occur in pairs like $\hat{c}^\dag_i\otimes \hat{c}^\dag_i$. Two such terms always commute, $[\hat{c}^\dag_i\otimes \hat{c}^\dag_i,\hat{c}^\dag_j\otimes \hat{c}^\dag_j]=0$ $\forall_{i,j}$.

\subsection{Bond dimensions}\label{sec:CE_dims}
Let us shortly discuss the bond dimensions required for matrix product representations of the canonical infinite-temperature states. This is most transparently done in an operator-space representation. In the following, we will consider a bipartition of the lattice $\mc{L}\cup\mc{R}$ into a left part $\mc{L}:=[1,i]$ and a right part $\mc{R}:=[i+1,L]$. 

The required matrix dimensions $D_i$ of an accurate matrix product representation \eqref{eq:MPP} of $\dm^\ce_{0,Q}$, or the corresponding purification $|\dmp^\ce_{0,Q}\ket$, can be determined by writing it as a sum of tensor product operators;
\begin{equation}\label{eq:CEinf_decomp}
	\dm^\ce_{0,Q} \equiv \id_Q = \sum_{Q'} \id^{\mc{L}}_{Q'} \otimes \id^{\mc{R}}_{Q-Q'},
\end{equation}
where $\id^{\mc{L}}_{Q'}$ is the identity in the quantum number $Q'$ subspace of subsystem $\mc{L}$. Analogously, $\id^{\mc{R}}_{Q''}$ is the identity in the quantum number $Q''$ subspace of subsystem $\mc{R}$.

\begin{figure}[t]
\includegraphics[width = \columnwidth]{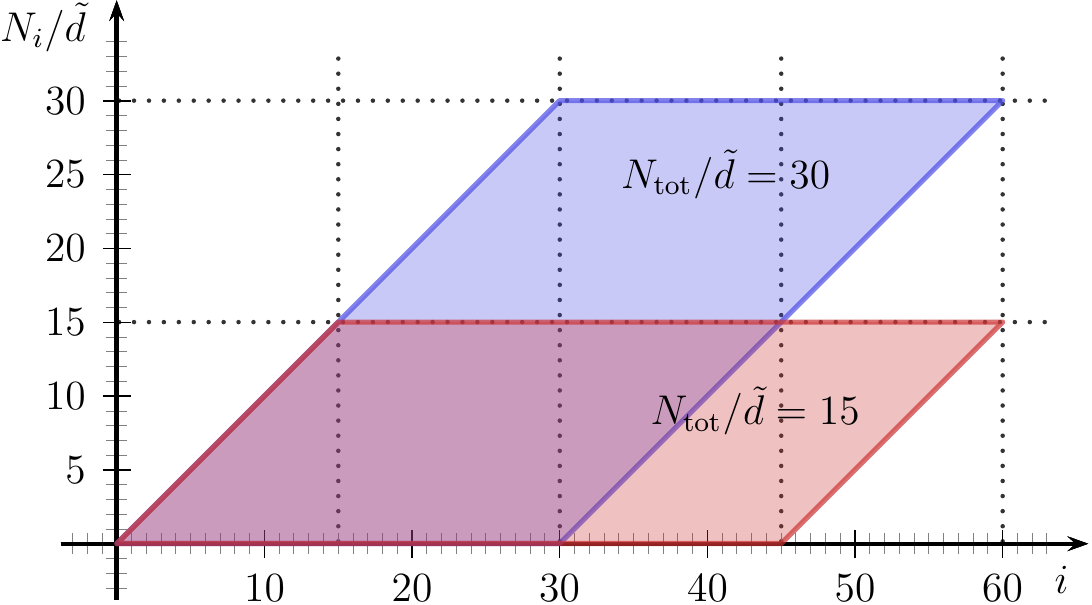}
\caption{\label{fig:ce_Nrange}Ranges \eqref{eq:N_constraints} of possible left-block quantum numbers $N_i=\sum_{j=1}^{i} n_i$ for a bosonic or fermionic system with $N_\tot$ particles or a spin-$s$ system with $N_\tot:=S^z_\tot+sL$. The system has $L=60$ sites and the diagram displays quantum number ranges for fillings $N_\tot=30\tilde{d}$ and $N_\tot=15\tilde{d}$, where $\tilde{d}=d-1$ is the maximum possible or considered number of particles per site -- in particular, $\tilde{d}=2s$ for spin-$s$. In matrix product representations \eqref{eq:MPP_CE_beta0} of the infinite-temperature canonical ensembles, every quantum number sector has dimension one such that bond dimensions are $D_i=N^\text{max}_i-N^\text{min}_i+1$. For fixed filling $N_\tot/L$, the maximum bond dimension is hence proportional to $L$.}
\end{figure}

For a bosonic or fermionic system with $Q=N$ particles (or a spin-$s$ system with $Q=N=S^z_\tot+sL$) and a bipartition at a bond $(i,i+1)$ sufficiently far away from the system boundaries, the sum in the decomposition \eqref{eq:CEinf_decomp} would contain $N+1$ orthogonal terms, implying that we need dimensions $D_i=N+1$ for the matrix product representations \eqref{eq:MPP} of $\dm^\ce_{0,N}$ or the corresponding purification $|\dmp^\ce_{0,N}\ket$. For a cut at a bond $(i,i+1)$ close to the left boundary, with $(d-1)i<N$, the sum would only contain $(d-1)i+1$ terms (the left subsystem can contain $0$ to $(d-1)i$ particles), and a bond dimension $D_i=(d-1)i+1<N+1$ would hence suffice at such bonds. See Fig.~\ref{fig:ce_Nrange}. In essence, for an exact matrix product representation of canonical infinite-temperature states, the maximum bond dimension $\max_i D_i$ increases linearly in the system size if we keep the particle number density $N/L$ fixed. In contrast, grand-canonical infinite-temperature states \eqref{eq:GCE_beta0} are product states such that $D_i=1$ $\forall_i$.

If one exploits symmetries as explained in section~\ref{sec:CE}, the computation costs for the canonical ensemble at high temperatures are actually lower than what one might expect from the growth of the MPS bond dimensions $D_i$. According to the decomposition \eqref{eq:CEinf_decomp} of the infinite-temperature state, every symmetry sector has dimension $1$ such that the matrices $A^{n_i,n_i'}_i$ in the matrix product representation consist of $D_i$ blocks of size $1\times 1$. Hence, at high temperatures, computation costs do not scale as $D_i^3$ but are instead linear in $D_i$.

\subsection{Exact matrix product representation}\label{sec:CE_beta0_exact}
Based on the decomposition
\begin{align*}
	\dm^\ce_{0,Q} = \sum_{q_1} \id^{\{1\}}_{q_1} \otimes \id^{\{2,\dotsc,L\}}_{Q-q_1}
		=\dotso = \sum_{\vec{q},\sum_i q_i=Q} \bigotimes_{i=1}^L \id^{\{i\}}_{q_i},
\end{align*}
we can give an explicit matrix product representation of the canonical infinite-temperature state. Here, $\id^{\mc{G}}_{Q'}$ denotes the identity in the quantum number $Q'$ subspace of the subsystem $\mc{G}$. A valid matrix product representation is given by
\begin{subequations}\label{eq:MPP_CE_beta0}
\begin{gather}
	\dm^\ce_{0,Q}=\sum_{\vn,\vn'} A^{n_1,n_1'}_1 A^{n_2,n_2'}_2\dotsb A^{n_L,n_L'}_L|\vn\ket\bra\vn'|,\\
	\text{with}\quad
	[A^{n_i,n_i'}_i]_{Q',Q''}=\delta_{n_i,n_i'}\delta_{Q'+q(n_i),Q''}.\label{eq:MPP_CE_beta0_mat}
\end{gather}
\end{subequations}
Here, we have used $Q'$ and $Q''$ as bond indices, which label at the same time the total quantum numbers for subsystems $\{1,\dots,i-1\}$ and $\{1,\dots,i\}$, respectively. They are hence eigenvalues of $\hQ'=\sum_{j=1}^{i-1}\hq_j$ and $\hQ''=\sum_{j=1}^i\hq_j$. Using the matrices of the MPDO \eqref{eq:MPP_CE_beta0} in Eq.~\eqref{eq:MPP} gives the infinite-temperature MPP $|\dmp^\ce_{0,Q}\ket$. The tuples $(q,\tilde{q})$ of quantum numbers, assigned to bond indices of the MPP as discussed in section~\ref{sec:CE}, are then simply $q(Q')=\tilde{q}(Q')=Q'$ and $q(Q'')=\tilde{q}(Q'')=Q''$.
The possible ranges of the indices and hence the matrix dimensions are easy to determine. As an example, consider again a bosonic or fermionic system with $Q=N$ particles or a spin-$s$ system with $Q=N=S^z_\tot+sL$. In this case, $Q'=N'$ and $Q''=N''$ denote subsystem particle numbers, obeying the constraints
\begin{alignat}{5}
	&\max\big(0,N-(L-i+1)\tilde{d}\big)&&\leq N' &&\leq \min\big(N,(i-1)\tilde{d}\big),\nonumber\\
	&\max\big(0,N-(L-i)\tilde{d}\big)  &&\leq N''&&\leq \min\big(N,i\tilde{d}\big), \label{eq:N_constraints}
\end{alignat}
where $\tilde{d}:=d-1$ is the maximum (allowed) number of particles per site. See Fig.~\ref{fig:ce_Nrange}. The MPP of the grand-canonical infinite-temperature state, given in section~\ref{sec:GCE}, can be considered as a special case of Eq.~\eqref{eq:MPP_CE_beta0} through assigning to all states the same trivial quantum number $q(n_i)=Q=Q'=Q''=0$, such that all bond dimensions are $1$.

In the formulation above, the approach is generically applicable. As a specific example, consider the Fermi-Hubbard model which has local basis states $|n_{i,\ua},n_{i,\da}\ket$ with occupation numbers $n_{i,\sigma}\in\{0,1\}$. Assume the ensemble where the total number of electrons $\hat{N}=\sum_{i,\sigma}\hat{n}_{i,\sigma}$ is fixed to $N$ and the magnetization expectation value $\bra \hS^z_\tot\ket=\frac{1}{2}\sum_i \bra \hat{n}_{i,\ua}-\hat{n}_{i,\da}\ket$ is fixed by a magnetic field. The $A$-tensors of the corresponding infinite-temperature state are then $[A^{(n_\ua,n_\da),(n'_\ua,n'_\da)}_i]_{N',N''}=\delta_{n_\ua,n_\ua'}\delta_{n_\da,n_\da'}\delta_{N'+n_\ua+n_\da,N''}$, where $N'$ and $N''$ take integer values as specified in Eq.~\eqref{eq:N_constraints} with $\tilde{d}=2$.

\subsection{Variational scheme using entangler Hamiltonians} \label{sec:entangler}
Earlier, it has been suggested to obtain the infinite-temperature purifications $|\dmp^\gce_{0,\alpha}\ket\in\H\otimes\H$ and $|\dmp^\ce_{0,Q}\ket\in\H_Q\otimes\H_Q$ [Eqs.~\eqref{eq:GCE_beta0} and \eqref{eq:CE_beta0}] by DMRG ground-state computations for so-called entangler Hamiltonians \cite{Feiguin2010-81,Nocera2016-93}. Let us shortly explain this approach and point out why, especially for the canonical ensembles, it should be less favorable than the more direct generating procedures described in sections~\ref{sec:CE_beta0_MPO} and \ref{sec:CE_beta0_exact}.

For grand-canonical infinite-temperature ensembles \eqref{eq:GCE_beta0} a variational computation is not really necessary. As described in section~\ref{sec:GCE}, there is an explicit simple matrix product representation with bond dimension $D_i=1$ $\forall_i$. If, however, for some practical reason, one needs to prepare this infinite-temperature purification by a variational computation, this can be done by computing $|\dmp^\gce_{0,\alpha}\ket\in\H\otimes\H$ as the zero-energy ground state of the Hamiltonian
\begin{equation*}
	\hat{\H}^\gce= \sum_i \hat{\mc{P}}^\gce_i,
\end{equation*}
where $\id-\hat{\mc{P}}^\gce_i$ projects sites $i$ of the primary system and the auxiliary system onto the maximally entangled state $\sum_{n_i}|n_i\ket\otimes|n_i\ket_\aux$ occurring in Eq.~\eqref{eq:GCE_beta0}.
\begin{figure}[t]
\includegraphics[width = \columnwidth]{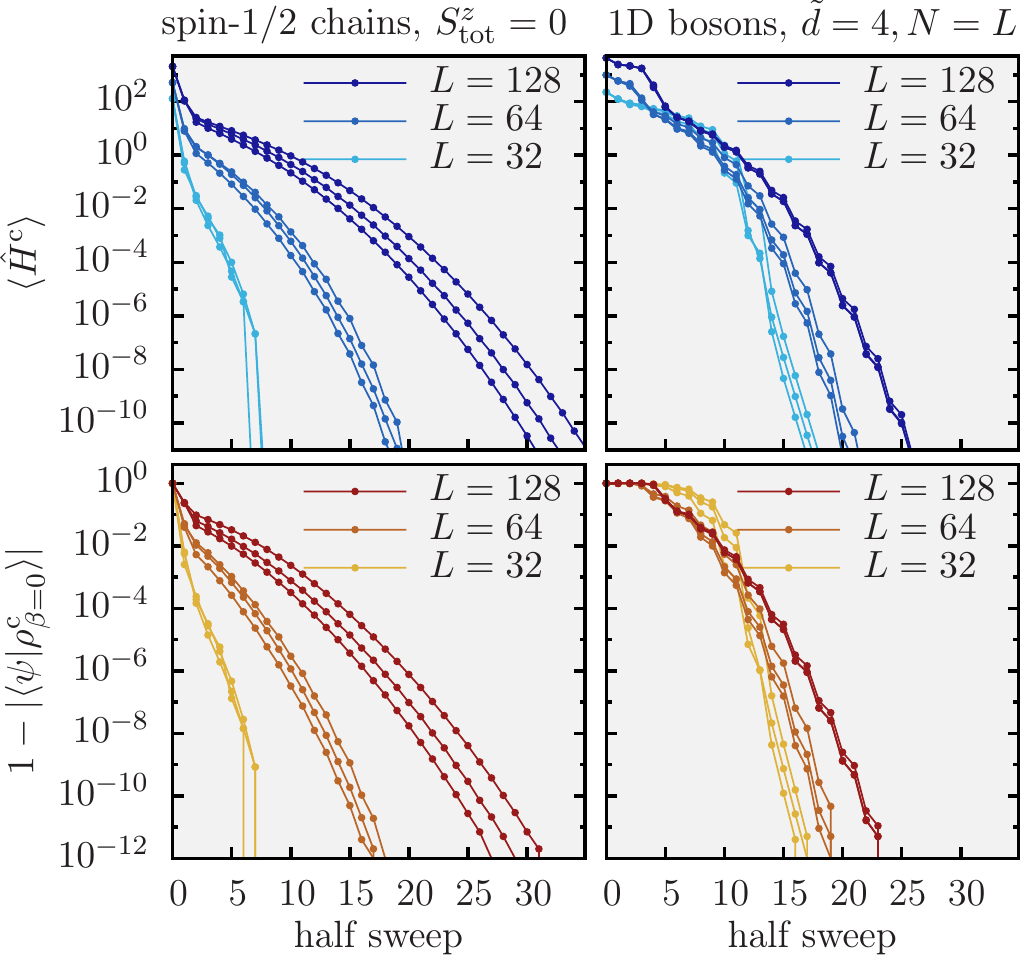}
\caption{\label{fig:ce_Hentang}Variational approach to construct MPPs of canonical infinite-temperature states \eqref{eq:CE_beta0} in a DMRG ground state computation with a so-called entangler Hamiltonian \eqref{eq:Hentang_ce}. The figure shows the convergence to the closely related equal-weight superposition $|\tilde\dmp^\ce_{0,Q}\ket$ in Eq.~\eqref{eq:equalWeight}, which is the zero-energy ground state of $\hH^\ce$ [Eq.~\eqref{eq:Hentang_ceDiag}]. Top: energy expectation value $\bra\hH^\ce\ket$. Bottom: fidelity measure $1-|\bra\psi|\tilde\dmp^\ce_{0,Q}\ket|$. Left: spin-$1/2$ chains of different lengths with magnetization $S^z_\tot=0$. Right: bosons on 1D lattices with a maximum of $\tilde{d}=4$ bosons per site and unit filling $N/L=1$. Different curves for the same system size $L$ refer to different initial random MPSs for the variational computation. Bond dimensions have been fixed to those of the corresponding exact MPP representations given in section~\ref{sec:CE_beta0_exact}.}
\end{figure}

Canonical infinite-temperature states \eqref{eq:CE_beta0} are less trivial. In sections~\ref{sec:CE_beta0_MPO} and \ref{sec:CE_beta0_exact}, we have discussed two direct approaches to generate $|\dmp^\ce_{0,Q}\ket\in\H_Q\otimes\H_Q$.
As suggested in Refs.~\cite{Feiguin2010-81,Nocera2016-93}, an alternative procedure is to compute $|\dmp^\ce_{0,Q}\ket$ variationally. In a somewhat modified and generalized form, applicable to arbitrary systems, this approach consists in computing the ground state of the Hamiltonian
\begin{equation}\label{eq:Hentang_ce}
	\hat{\H}^\ce:= \sum_{i<j} \hat{\mc{P}}^\ce_{i,j}
	\quad\text{with}\quad 
	\hat{\mc{P}}^\ce_{i,j}=\id-\hat{\mc{X}}_{i,j}
\end{equation}
in $\H_Q\otimes\H_Q$.
The positive semidefinite operators $\hat{\mc{P}}^\ce_{i,j}\succeq 0$ act on sites $i$ and $j$ of both the primary and the auxiliary systems. Specifically, $\hat{\mc{X}}_{i,j}$ denotes what we call a \emph{shuffle operator},
\begin{equation*}\label{eq:Shuffle}
	\hat{\mc{X}}_{i,j} := \hspace{-6ex}\sum_{\substack{n_i,n_j,n'_i,n'_j\\ q(n_i)+q(n_j)=q(n'_i)+q(n'_j)}} \hspace{-6ex}\frac{|n_i',n_j'\ket\bra n_i,n_j|\otimes(|n_i',n_j'\ket\bra n_i,n_j|)_\aux }{V_{q(n_i)+q(n_j)}},
\end{equation*}
where $V_k:=\sum_{n_i,n_j}\delta_{k,q(n_i)+q(n_j)}$ is the dimension of the two-site subspace with quantum number $k$. As will be shown below, $|\dmp^\ce_{0,Q}\ket$ is the unique ground state of the Hamiltonian \eqref{eq:Hentang_ce} in the diagonal subspace
\begin{equation*}
	\H^\diag_Q :=\Span\big\{|\vn\ket\otimes|\vn\ket\,|\sum_{i=1}^L q(n_i)=Q\big\} \,\subset\, \H_Q\otimes\H_Q
\end{equation*}
However, besides the complication of long-range interactions and comparatively strong entanglement in $|\dmp^\ce_{0,Q}\ket$ which can cause a slow and computationally costly convergence, there is another possible pitfall. If the variational MPS optimization to compute the ground state of $\hat{\H}^\ce$ is started with a state that happens to have no component in the diagonal subspace $\H^\diag_Q$, the optimization would not converge to $|\dmp^\ce_{0,Q}\ket$, but to a state in the orthogonal complement of $\H^\diag_Q$.

The latter problem can in fact be resolved by restricting the optimization to the diagonal subspace as follows. Instead of computing the ground state $|\dmp^\ce_{0,Q}\ket$ of $\hat{\H}^\ce$ in $\H_Q\otimes\H_Q$, we can compute the closely related equal-weight superposition
\begin{equation}\label{eq:equalWeight}
	|\tilde\dmp^\ce_{0,Q}\ket:= \sum_{\vn,\sum_i q(n_i)=Q} |\vn\ket \,\in\,\H_Q
\end{equation}
($\H_Q$ is isomorphic to $\H^\diag_Q$), which, in the following, will be shown to be the unique ground state of
\begin{equation}\label{eq:Hentang_ceDiag}
	\hH^\ce:= \sum_{i< j} \hat{P}^\ce_{i,j}
	\quad\text{with}\quad 
	\hat{P}^\ce_{i,j}=\id-\hX_{i,j}.
\end{equation}
This is the restriction of $\hat{\H}^\ce$ to the diagonal subspace $\H^\diag_Q$ with shuffle operators
\begin{equation*}
	\hX_{i,j} := \hspace{-2ex}\sum_{\substack{n_i,n_j,n'_i,n'_j\\ q(n_i)+q(n_j)=q(n'_i)+q(n'_j)}} \hspace{-4ex}\frac{|n_i',n_j'\ket\bra n_i,n_j|}{V_{q(n_i)+q(n_j)}}.
\end{equation*}
After computing the ground state $|\tilde\dmp^\ce_{0,Q}\ket$ of this Hamiltonian in MPS form $|\tilde\dmp^\ce_{0,Q}\ket=\sum_{\vn}\tilde{A}^{n_1}_1 \tilde{A}^{n_2}_2\dotsb \tilde{A}^{n_L}_L|\vn\ket$, the MPP $|\dmp^\ce_{0,Q}\ket$ in the form \eqref{eq:MPP} is obtained by defining $A^{n,n'}_i:=\delta_{n,n'}\tilde{A}^{n}_i$ $\forall_{i,n,n'}$.

Let us discuss why the equal-weight superposition $|\tilde\dmp^\ce_{0,Q}\ket$ is indeed the unique ground state of $\hH^\ce$ in $\H_Q$. As $\hH^\ce$ is a sum of positive semidefinite operators $\hat{P}^\ce_{i,j}\succeq 0$ \footnote{The shuffle operators $\hX_{i,j}$ are block-diagonal. Each block corresponds to a fixed quantum number $k=q(n_i)+q(n_j)$. In the two-site basis $\{|n_i,n_j\ket\,|\,q(n_i)+q(n_j)=k\}$ the block matrix is $J_{V_k}/V_k$, where $J_m$ denotes the $m\times m$ all-ones matrix. $J_m/m$ has one eigenvalue 1 and all others are zero. Consequently, the operators $\hat{P}^\ce_{i,j}=\id-\hX_{i,j}$ of Eq.~\eqref{eq:Hentang_ceDiag} have, in each quantum number sector of the two-site Hilbert space, one eigenvalue 0 and all others 1. And thus, $\hat{P}^\ce_{i,j}\succeq 0$, and also $\hat{\mc{P}}^\ce_{i,j}\succeq 0$ in Eq.~\eqref{eq:Hentang_ce}.}, its smallest possible eigenvalue is zero, i.e., $\hH^\ce\succeq 0$. Additionally, $\hH^\ce|\tilde\dmp^\ce_{0,Q}\ket=0$, such that $|\tilde\dmp^\ce_{0,Q}\ket$ is \emph{a} ground state of $\hH^\ce$. Furthermore, it is \emph{the} unique ground state according to the Perron-Frobenius theorem, applied to $\sum_{i< j} \hX_{i,j}$ in the basis $\{|\vn\ket\}$, which is a Hermitian irreducible non-negative matrix. 

Figure~\ref{fig:ce_Hentang} shows the convergence behavior of the variational computation of $|\tilde\dmp^\ce_{0,Q}\ket$. For the spin-$1/2$ chains, one sees that the required number of DMRG sweeps increases substantially with the system size $L$. For the bosonic systems, the required number of sweeps does not change as strongly with the system size.

\section{Computing quantum number distributions} \label{sec:MPDO_QN-distr}
The exact matrix product representations \eqref{eq:MPP_CE_beta0} of canonical infinite-temperature states $\dm^\ce_{\beta=0,Q}=\id_Q$ allow us to compute the weights $p_Q$ of quantum numbers $Q$ in any other MPDO $\dm$. With $\delta_{\hQ,Q}\equiv\dm^\ce_{0,Q}$, the probabilities $p_Q$ can be obtained by evaluating Hilbert-Schmidt inner products
\begin{equation}\label{eq:probabilityDistr}
	 p_Q = \bra\delta_{\hQ,Q}\ket_\dm = \Tr(\dm\,\dm^\ce_{0,Q}).
\end{equation}

As an example, let us consider the grand-canonical ensemble \eqref{eq:GCE}
\begin{equation*}
	\dm=\frac{\dm^\gce_{\beta,\alpha}}{Z}
	 \equiv \frac{e^{-\beta(\hH+\alpha\hQ)}}{Z}
	 = \sum_Q p_Q\frac{e^{-\beta\hH_Q}}{Z_Q}.
\end{equation*}
As described above, DMRG computations are often formulated in terms of purifications instead of MPDOs. With MPPs $|\dmp^\gce_{0,\alpha}\ket\equiv |\id\ket$, $|\dmp^\gce_{\beta,\alpha}\ket$, and $|\dmp^\ce_{0,Q}\ket\equiv |\id_Q\ket$, the quantum number probability distribution is then given by 
\begin{equation*}
	 p_Q = \frac{\bra\dm^\ce_{0,Q}|\dmp^\gce_{\beta,\alpha}\ket}{\bra\dmp^\gce_{0,\alpha}|\dmp^\gce_{\beta,\alpha}\ket}.
\end{equation*}
Here, we have used the grand-canonical infinite-temperature state to obtain the normalization factor $Z=\Tr\dm^\gce_{\beta,\alpha} =\bra\dmp^\gce_{0,\alpha}|\dmp^\gce_{\beta,\alpha}\ket$. Note that, whereas one employs the purifications $|\dmp^\gce_{\beta/2,\alpha}\ket$ for the evaluation of expectation values at inverse temperature $\beta$ [Eq.~\eqref{eq:GCE_expect}], one needs to go twice as far in imaginary time and use $|\dmp^\gce_{\beta,\alpha}\ket$ for the evaluation of probabilities $p_Q$ at inverse temperature $\beta$.

\section{An example} \label{sec:example}
\begin{figure}[b]
\includegraphics[width = 0.8\columnwidth]{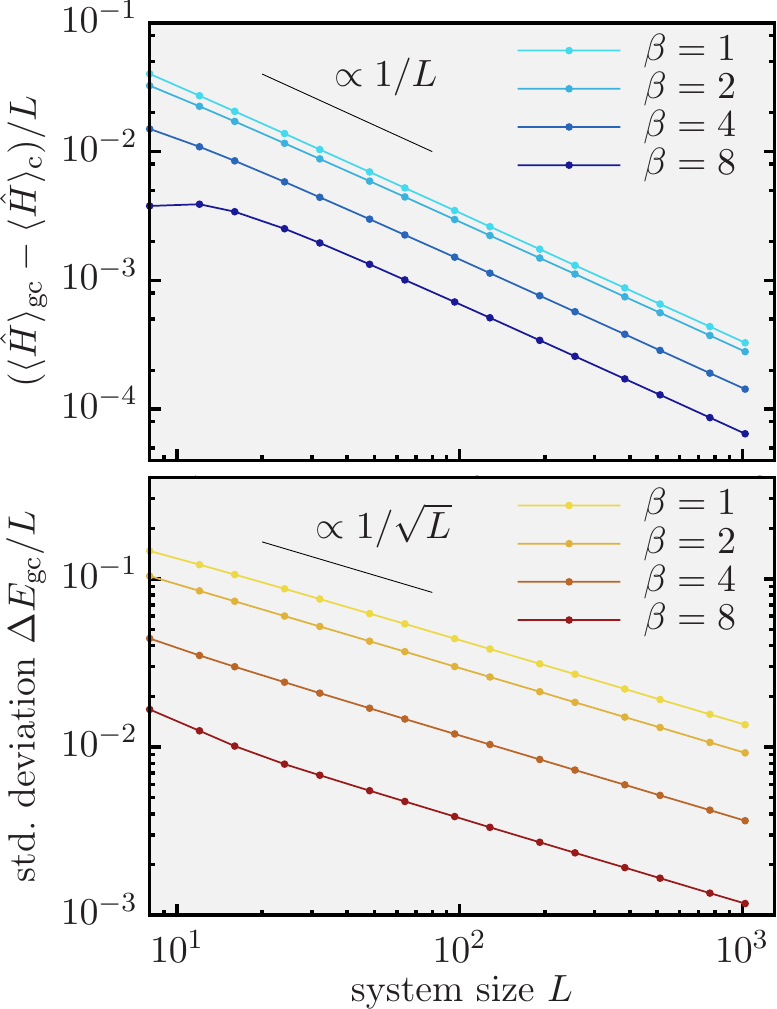}
\caption{\label{fig:ensembleEquivalence}MPP simulations of the canonical and grand-canonical ensembles for zero magnetization, which are equivalent in the thermodynamic limit. For the antiferromagnetic spin-$1/2$ Heisenberg chain \eqref{eq:H_XXZ}, the upper panel shows the difference of the energy densities for the two ensembles that decays as $1/L$ with the system size $L$. The lower panel shows the standard deviation of the energy density in the grand-canonical ensemble. As is known from thermodynamics, it decays as $1/\sqrt{L}$. }
\end{figure}
Let us exemplify the described matrix product techniques for antiferromagnetic spin-$1/2$ Heisenberg chains
\begin{equation}\label{eq:H_XXZ}
	\hH = \sum_{i=1}^{L-1} \hvS_i\cdot\hvS_{i+1},
\end{equation}
where $\hvS_i=(\hS^x_i,\hS^y_i,\hS^z_i)$. We consider zero magnetization, i.e., magnetic field $h^z=0$ for the grand-canonical ensemble and $\hS^z_\tot$-eigenvalue $Q=M=0$ for the canonical ensemble. We do quasi-exact MPP/MPDO simulations of both ensembles with imaginary-time step $\Delta\tau=1/16$ and a truncation threshold $\epsP=10^{-14}$, which fixes the matrix product bond dimensions dynamically as detailed in Refs.~\cite{Barthel2013-15,Binder2015-92}.

The upper panel of Fig.~\ref{fig:ensembleEquivalence} shows the difference of the energy densities for the two ensembles. It decays as $1/L$, indicating that the two ensembles are equivalent in the thermodynamic limit. The lower panel shows the standard deviation of the energy density in the grand-canonical ensemble. It decays as $1/\sqrt{L}$. This $1/\sqrt{L}$ decay is well-known from the theory of thermodynamics. It follows from the fact that the energy is extensive and that its derivative with respect to the inverse temperature $\beta$, which is intensive, yields the energy variance.
\begin{align}
	\frac{\partial}{\partial\beta}\bra \hH\ket_\gce &= \mc{O}(L)\quad \text{and} \nonumber\\
	-\frac{\partial}{\partial\beta}\bra \hH\ket_\gce&= \bra \hH^2\ket_\gce - \bra \hH\ket_\gce^2= \Delta E^2_\gce, \label{eq:DeltaE}
\end{align}
such that $\Delta E_\gce/L=\mc{O}(1/\sqrt{L})$ \footnote{For nonzero magnetic field, the extensiveness of $\Delta E^2_\gce$ follows, e.g., from $\big(\frac{\partial}{\partial\beta}+\frac{h}{\beta}\frac{\partial}{\partial h}\big)\bra \hH\ket_\gce=\bra \hH\ket_\gce^2-\bra \hH^2\ket_\gce$.}.
\begin{figure}[t]
\includegraphics[width = \columnwidth]{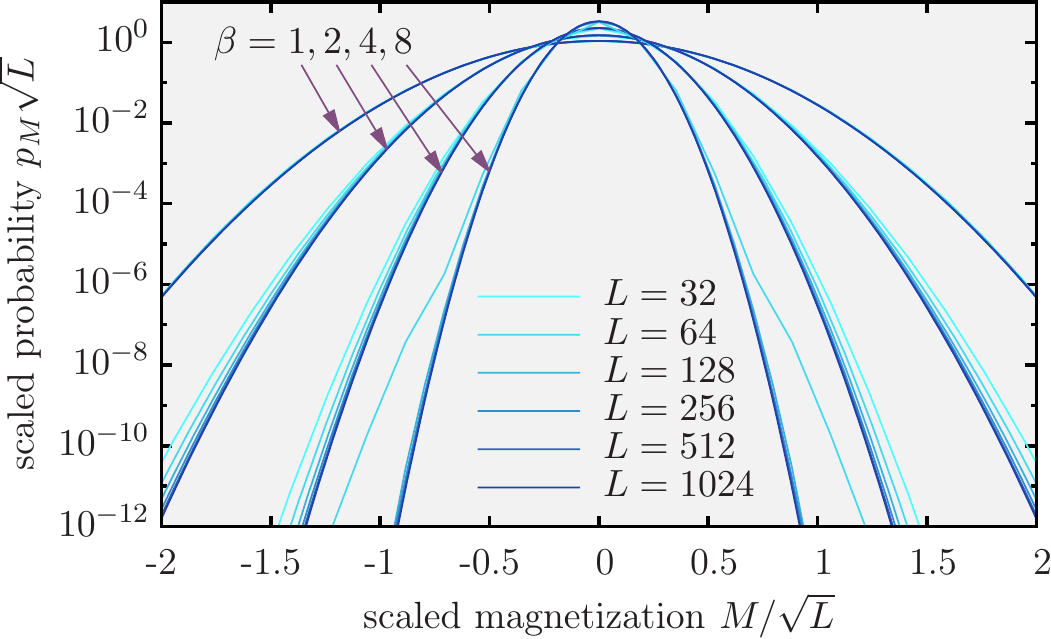}
\caption{\label{fig:gce_SzDistribution}Probability distribution $p_M$ [Eq.~\eqref{eq:probabilityDistr}] of the total magnetization $M$ in the grand-canonical ensemble $e^{-\beta\hH}/Z=\sum_M p_M e^{-\beta\hH_M}/Z_M$ for antiferromagnetic spin-$1/2$ Heisenberg chains \eqref{eq:H_XXZ} without magnetic field, computed with MPDO simulations as described in section~\ref{sec:MPDO_QN-distr}. The properly rescaled distribution $(M,p_M)\to(M/\sqrt{L},p_M/\sqrt{L})$ converges in the thermodynamic limit $L\to\infty$.}
\end{figure}

\begin{figure}[t]
\includegraphics[width = \columnwidth]{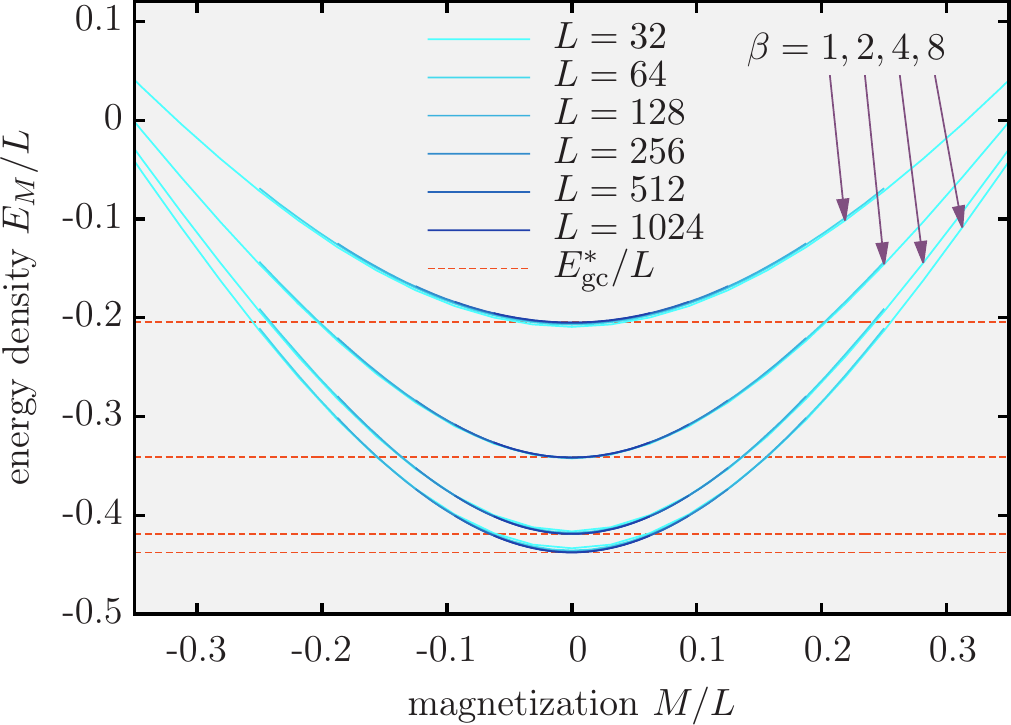}
\caption{\label{fig:ce_energy}Energy densities $E_M/L$ of canonical ensembles $e^{-\beta\hH_M}/Z_M$ for antiferromagnetic spin-$1/2$ Heisenberg chains \eqref{eq:H_XXZ}, computed with MPP simulations. The energy density $E_M/L$ as a function of the magnetization $M/L$ converges in the thermodynamic limit $L\to\infty$. The energy densities for the different temperatures have minima at $M=0$ as this is the ground-state sector.
Due to the equivalency of ensembles, $\lim_{L\to\infty}E_{M=0}/L$ coincides with the energy density $\lim_{L\to\infty}E_\gce/L$ of the grand-canonical ensemble $e^{-\beta\hH}/Z$ for zero magnetization $\bra\hS^z_\tot\ket_\gce=0$.}
\end{figure}

The decay of the energy density difference $\sim 1/L$ shown in the upper panel of Fig.~\ref{fig:ensembleEquivalence} can be explained as follows. According to thermodynamics, the standard deviation of the magnetization $\Delta M_\gce/L$ in the grand-canonical ensemble decays as $1/\sqrt{L}$, just like $\Delta E_\gce/L$ in Eq.~\eqref{eq:DeltaE}. Beyond that, the appropriately rescaled magnetization probability density converges in the thermodynamic limit such that $p_M\to g(M/\sqrt{L})/\sqrt{L}$. Using the technique described in section~\ref{sec:MPDO_QN-distr}, we have computed the magnetization probabilities $p_M$ in the grand-canonical ensemble. The results are displayed in Fig.~\ref{fig:gce_SzDistribution}. The scaling function $g$ is approximately of Gaussian shape for the considered temperatures. We have also computed the energies $E_M=\Tr(\hH\,\dm^\ce_{\beta,M})/Z_M$ in canonical ensembles for total magnetizations $M$. As shown in Fig.~\ref{fig:ce_energy}, the energy density $E_M/L$ as a function of the magnetization $M/L$ converges with increasing $L$. Because the global ground state of the isotropic Heisenberg antiferromagnet \eqref{eq:H_XXZ} has zero magnetization, the energy density has a minimum at $M=0$ for the considered temperatures. Due to the global spin-flip symmetry, the scaling function is symmetric around $M=0$ and is found to be of the form $E_{M=0}/L+\kappa (M/L)^2$. Combining these observations, we find for the difference of the canonical and grand-canonical energy densities that
\begin{align*}
	&\frac{\bra\hH\ket_\gce-\bra\hH\ket_\ce}{L}
	 = \sum_M p_M\, \frac{E_M-E_{M=0}}{L}\\
	 &\quad\to \frac{\kappa}{\sqrt{L}} \int \mathrm{d}M\, g\left(\frac{M}{\sqrt{L}}\right) \left(\frac{M}{L}\right)^2
	 = \frac{\kappa}{L} \int \mathrm{d}x\, g(x) x^2
\end{align*}
which explains the $1/L$ scaling in the upper panel of Fig.~\ref{fig:ensembleEquivalence}.

The same $1/L$ scaling of the energy density difference,
\begin{equation*}
	(\bra \hH\ket_\gce - \bra \hH\ket_\ce)/L\sim 1/L,
\end{equation*}
occurs for nonzero magnetizations, i.e., if we compare the canonical ensemble $e^{-\beta\hH_{\tilde{M}}}/Z_{\tilde{M}}$ for total magnetization $\tilde{M}$ to the grand-canonical ensemble with magnetic field $h^z$ chosen such that $\bra \hS^z_\tot\ket_\gce=\tilde{M}$. In that case, $E_M-h^z M$ has its minimum at $M=\tilde{M}$ and the probability distribution converges like $p_M\to \tilde{g}\big((M-\tilde{M})/\sqrt{L}\big)/\sqrt{L}$.

\section{Conclusions}
It is straightforward to compute an MPP/MPDO of a grand-canonical ensemble, as the infinite-temperature state is simply a product state and has hence a matrix product representation with bond dimension 1 (Sec.~\ref{sec:GCE}). For systems with symmetries it can be computationally beneficial to work instead with canonical ensembles. While, for a system with a conserved quantity $\hQ$, grand-canonical ensembles obey $[\hQ,\dm^\gce_{\beta,\alpha}]=0$, canonical ensembles obey $\dm^\ce_{\beta,Q}\hQ=\hQ\dm^\ce_{\beta,Q}=Q\dm^\ce_{\beta,Q}$. This can be used to reduce computation costs substantially (Sec.~\ref{sec:CE}). Sometimes, canonical ensembles also better reflect experimental conditions (e.g., fixed particle number in experiments with ultracold atoms). To compute matrix product representations of canonical ensembles is somewhat more complicated as the corresponding infinite-temperature states are nontrivial. We have described and discussed different techniques to generate these states in matrix product form (Sec.~\ref{sec:CE_beta0}). The cleanest way is certainly to directly use the exact matrix product representation \eqref{eq:MPP_CE_beta0}. If this is not feasible for practical reasons, one can alternatively generate canonical infinite-temperature states by applying MPOs multiple times to vacuum-type states (Sec.~\ref{sec:CE_beta0_MPO}) or by doing variational computations with entangler Hamiltonians (Sec.~\ref{sec:entangler}). In the latter case, one should either work in the diagonal subspace or make at least sure that the initial state for the variation has components in the diagonal subspace.

Another useful application of the constructed canonical infinite-temperature states in matrix product form is to study probability distributions of global quantum numbers in mixed states (MPDOs). To this purpose one needs to evaluate corresponding Hilbert-Schmidt inner products (Sec.~\ref{sec:MPDO_QN-distr}).

\textbf{Acknowledgments. --}
Discussions with M.\ Binder, G.\ Roux, G.\ Alvarez, and A.\ Nocera and a careful reading of the manuscript by M.\ Binder are gratefully acknowledged.

\bibliographystyle{prsty}

\end{document}